# Fatigue and retention in the growth window of ferroelectric Hf$_{0.5}$Zr$_{0.5}$O$_2$ thin films

Jike lyu, Ignasi Fina,* Florencio Sánchez*

Institut de Ciència de Materials de Barcelona (ICMAB-CSIC), Campus UAB, Bellaterra 08193, Barcelona, Spain

**Abstract**

The growth window of epitaxial Hf$_{0.5}$Zr$_{0.5}$O$_2$ is established taking into account the main ferroelectric properties that films have to present simultaneously: high remanent polarization, low fatigue and long retention. Defects in the film and imprint field depend on deposition temperature and oxygen pressure, with an impact on fatigue and retention, respectively. Fatigue increases with substrate temperature and pressure, and retention is short if low temperature is used. The growth window of epitaxial stabilization of ferroelectric Hf$_{0.5}$Zr$_{0.5}$O$_2$ is narrower when all major ferroelectric properties (remanence, endurance and retention) are considered, but deposition temperature and pressure ranges are still sufficiently wide.

* Corresponding authors: ifina@icmab.es, fsanchez@icmab.es

Ferroelectric $HfO_2$ films can be deposited using CMOS-compatible conditions, and have a large remanent polarization and retention for more than 10 years, even when the thickness is less than 10 nm.[1,2] These characteristics make ferroelectric $HfO_2$ more suitable for applications than conventional ferroelectric perovskites. However, the endurance of ferroelectric $HfO_2$ is significantly less than that of ferroelectric perovskites. Endurance is critical for memory devices and therefore, improvement of this property in $HfO_2$-based ferroelectrics is extremely important. Efforts to achieve this goal have allowed endurance of up to $10^{11}$ cycles to be achieved.[3,4] Further progress is challenging due to polarization - endurance[2] and retention - endurance[5] dilemmas. Recently, we have achieved high polarization, retention, and endurance in sub-5 nm epitaxial $Hf_{0.5}Zr_{0.5}O_2$ (HZO) films.[4] However, the epitaxial films exhibited significant fatigue, and although the capacitors remain operational after $10^{11}$ cycles, the remanent polarization is greatly reduced. Reducing fatigue while maintaining high polarization and long retention is an important goal. The effects of film thickness and deposition temperature and pressure on the ferroelectric polarization of epitaxial HZO were determined.[6] Films deposited at around 800 °C and 0.1 mbar showed the highest polarization.[6] Very recently, the impact of thickness on endurance and retention has been reported.[4] It is essential to determine the effects of the growth window (deposition temperature and oxygen pressure) on retention and fatigue. Here, we report on these properties in HZO films prepared in the growth window (deposition temperature and oxygen pressure) that allows epitaxial stabilization of the ferroelectric phase. It is shown that both deposition parameters have a great influence on retention and fatigue, and the growth window to have high retention and endurance is determined. The results are discussed considering the dielectric permittivity, and imprint and coercive fields of the films. The growth window of epitaxial HZO films is revised to include all main ferroelectric properties: remanent polarization, fatigue and retention.

HZO films of thickness (t) ~9 nm were grown on $SrTiO_3(001)$ (STO) substrates buffered with an $La_{2/3}Sr_{1/3}MnO_3$ (LSMO) electrode of thickness ~25 nm. The ferroelectric HZO film and the LSMO electrode were deposited in a single process by pulsed laser deposition (PLD) using a KrF excimer laser. LSMO electrodes were deposited at 700 °C under 0.1 mbar of oxygen. In one series of samples ($T_s$ series), HZO films were deposited at an oxygen pressure of 0.1 mbar and a variable substrate temperature $T_s$ from 650 to 825 °C. In a second series ($P_{O2}$ series), the temperature of the substrate was fixed (800 °C) and a variable oxygen pressure $P_{O2}$ in the range of 0.01 - 0.2 mbar was used. Immediately after HZO growth, the samples were cooled under 0.2 mbar of oxygen. The HZO/LSMO bilayers are epitaxial. Detailed structural characterization of the samples and additional experimental details are reported elsewhere.[6,7]

Circular top platinum electrodes, of diameter ≈20 µm and thickness 20 nm, were deposited by sputtering at room temperature through stencil masks for electrical characterization (Supplementary material, Fig. S1). Area of the electrodes was determined for each sample after electrodes deposition (Supplementary material, Fig. S2). Ferroelectric polarization loops were measured using an AixACCT TFAnalyser2000 platform at 1 kHz by the Dynamic Leakage Current Compensation (DLCC) procedure[8,9] at room temperature in top-bottom configuration. Fatigue was measured cycling the HZO films $10^7$ times at frequency of 100 kHz using 4.5 V bipolar square pulses and measuring polarization loops at 1 kHz. Retention was measured at room temperature poling the sample using triangular pulses of 5.5 V and 0.25 ms and determining the remanent polarization from the first polarization curve of the polarization loop measured at 1 kHz using the PUND protocol[9] after a delay time. Different voltage applied in retention and endurance does not mimic a real device operation; however, it serves here to better visualize the retention and endurance trends on growth conditions. Capacitance (C) loops were measured by using an impedance analyzer (HP4192LF, Agilent Co.) operated with an excitation voltage of 50 mV at 20 kHz. Dielectric permittivity (ε)-voltage loops were extracted from capacitance values using the C = εA/t relation, where A is the electrode area and t is the film thickness.

Figure 1a shows representative polarization loops during the endurance measurement of the $T_s$ = 650 °C film ($P_{O2}$ = 0.1 mbar). They were obtained from current hysteresis loops shown in Supplementary material, Fig. S3. Remanent polarization is $P_r$ = 9.9 µC/cm$^2$ in the pristine state. Polarization decreases to $P_r$ = 8.5 µC/cm$^2$ after $10^3$ cycles and it is reduced to $P_r$ = 4.4 µC/cm$^2$ after $10^7$ cycles, indicating fatigue. Fatigue is more evident in the $T_s$ = 825 °C film, in which $P_r$ decreases from 11.9 to 4.8 µC/cm$^2$ after $10^7$ cycles (Figure 1b). The effect of $T_s$ on fatigue is summarized in Figure 1c, where the dependence of the normalized remanent polarization with the number of cycles is plotted for all samples of the $T_s$ series. The graph shows that remanent polarization is more reduced as $T_s$ increases from $T_s$ = 650 °C (47% after $10^7$ cycles) to 825 °C (27% after $10^7$ cycles). Polarization loops were also measured after 4 and 8 cycles at ~5.5 and 4.5 V (Supplementary material, Fig. S4 and S5, respectively) to determine if there is wake-up effect. The two films deposited at lowest temperature, $T_s$ = 650 °C and 700 °C) show a tiny wake-up effect, with very small increase of polarization after 4 and 8 cycles (Supplementary material, Fig. S6). Wake-up effect is completely absent in the films deposited at higher temperature (and fixed thickness of 9 nm and deposition pressure of 0.1 mbar).

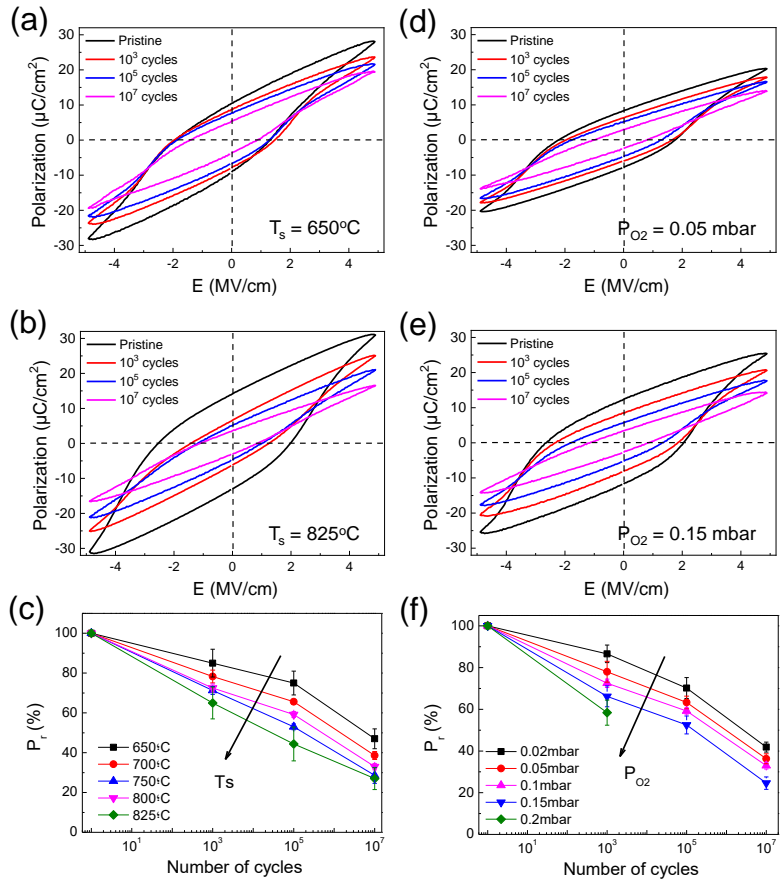

**Figure 1**: Ferroelectric hysteresis loops, measured in pristine state and after indicated number of cycles, for films deposited at $P_{O2}$ = 0.1 mbar and $T_s$ = 650 °C (a), 0.1 mbar and 825 °C (b), $P_{O2}$ = 0.05 mbar and $T_s$ = 800 °C (d), and 0.15 mbar and 800 °C (e). Variation with number of cycles of remanent polarization (positive and negative average values), normalized to its initial value, of films deposited at variable $T_s$ (c) and $P_{O2}$ (f).

The effect of $P_{O2}$ on fatigue is presented in Figures 1 d-f. Representative polarization loops of the $P_{O2}$ = 0.05 mbar and 0.15 mbar films ($T_s$ = 800 °C) are shown in Figures 1d and 1e, respectively. Fatigue is less in the low $P_{O2}$ film, in which the remanent polarization decreases less with cycling (from 8.3 μC/cm$^2$ in the pristine state to 2.7 μC/cm$^2$ after 10$^7$ cycles in the $P_{O2}$ = 0.05 mbar film) than in the high $P_{O2}$ film (from 12.0 to 3.0 μC/cm$^2$ in the $P_{O2}$ = 0.15 mbar film). Fatigue increases monotonically with $P_{O2}$ (Figure 1f), and moreover the $P_{O2}$ = 0.2 mbar film underwent hard breakdown after only 10$^3$ cycles. Fig. S7-9 in Supplementary material confirms that there is no wake-up effect in the investigated range of oxygen pressure, for fixed film thickness of around 9 nm thick and deposition temperature of 800 °C. Data in Supplementary material, Fig. S3 shows that the ferroelectric switching at very high $T_s$ and/or $P_{O2}$ evaluated after $10^7$ cycles is small but visible. The direct subtraction of the $P_r$ value from the ferroelectric switching peak shows that the overestimation done in the direct $P_r$ determination is around 2 μC/cm$^2$.

Retention measurements of the $T_s$ = 650 °C and 825 °C films ($P_{O2}$ = 0.1 mbar) are presented in Figures 2a and 2b, respectively. Measurements were made by positively poling the top Pt electrode (black square symbols) as well as negatively poling it (red circle symbols). The data are fitted to the equation $P_r = P_0 \, t_d^{-n}$ to quantify retention. To evaluate the retention in Figure 2c, it is represented by $P_r$ value extrapolated at 10 years, normalized to its initial value, as a function of $T_s$. There is a strong asymmetry in the $T_s$ = 650 and 700 °C film. The polarization is very stable if the sample is positively poled, with $P_r$(@10years) ≈ 80%. In contrast, $P_r$(@10years) is ≈ 10% for negative poling, signaling very short retention. The strong asymmetry limits the application of the $P_r = P_0 \, t_d^{-n}$ equation, because it does not take into account the role of imprint.[10,11] Imprint has an important impact on low $T_s$ samples retention due to its smaller coercivity, as discussed below. As the deposition temperature increases to $T_s$ = 750 °C, the small $P_r$(@10years) for negative poling increases to ≈ 40%, while for positive poling it decreases to ≈ 60%. Therefore, the asymmetry is considerably reduced. Films deposited at higher $T_s$ show similar retention characteristics. Figures 2d and 2e show retention graphs of the $P_{O2}$ = 0.05 and 0.15 mbar films ($T_s$ = 650 °C), respectively. For positive bias, $P_r$(@10years) is ≈ 60% in the $P_{O2}$ = 0.05 mbar film, and ≈ 65% in the $P_{O2}$ = 0.15 mbar film. The corresponding values for negative poling are ≈ 50% and ≈ 45%. Therefore, both films have a similar long retention, and non-significant asymmetry. The graph $P_r$(@10years) against $P_{O2}$ (Figure 2f) indicates that there is no significant influence of $P_{O2}$ on retention.

The imprint field ($E_{imp}$) and the coercive voltage ($V_c$) were determined from polarization loops. Its dependence on the deposition temperature and the oxygen pressure are represented in Figures 3a and 3b, respectively. $E_{imp}$ is almost constant in the temperature window, around 400 kV/cm, for all the samples of the $T_s$ series. $E_{imp}$ relies greatly on $P_{O2}$, increasing from a negligible value at 0.01 mbar to approximately 400 kV/cm at 0.1 mbar. The imprint field in polycrystalline doped-HfO$_2$ films is generally associated with oxygen vacancies defects.[12,13] $E_{imp}$ is very small in epitaxial films deposited at the lowest oxygen pressure, and its magnitude increases markedly with $P_{O2}$. This suggests that there is no simple relationship between oxygen stoichiometry and $E_{imp}$ in our films. The opposite dependence would be expected if $E_{imp}$ is due to oxygen vacancies. The results suggest that there is not great difference in oxygen vacancies in films grown by PLD in the $P_{O2}$ = 0.01-0.2 mbar range. Indeed, ferroelectric doped-HfO$_2$ films grown by sputtering have the largest polarization when pure argon atmosphere is used.[14,15] Also, formation of orthorhombic phase was demonstrated in undoped HfO$_2$ deposited under null or very low oxygen partial pressure.[16,17] In contrast, the orthorhombic phase is favored in epitaxial HZO films deposited by PLD at high oxygen pressure.[6] PLD is a highly energetic technique and $P_{O2}$, which reduces plasma energy, could influence the formation of charged defects. Probably, charged defects more complex than oxygen vacancies can be responsible for $E_{imp}$. The direction of $E_{imp}$ is from the upper to the lower electrode in all samples (see sketches in the insets of Figures 3a and 3b), which favors positive polarization while reducing the stability of negative polarization. The observed anisotropy in retention, with a positive polarization more stable than the negative, is explained by the direction of $E_{imp}$. On the other hand, the magnitude of $V_c$ is also expected to affect retention. $V_c$ increases with $T_S$ from 1.9 to 2.9 V (Figure 3a), and with $P_{O2}$ from 0.9 V to 2.8 V (Figure 3b). The summary shown in Figures 2c and 2f indicates similar retention characteristics in all films except those deposited at low $T_s$ (650 and 700 °C), which have high asymmetry. The distinctive characteristics of these films with respect to the others, is that they present a combination of low $V_c$ and high $E_{imp}$, which explains why the imprint field has a greater effect in retention (Figure 2c). Instead, when increasing $P_{O2}$ both magnitudes ($E_{imp}$ and $V_c$) increase with no important impact on retention (Figure 2f).

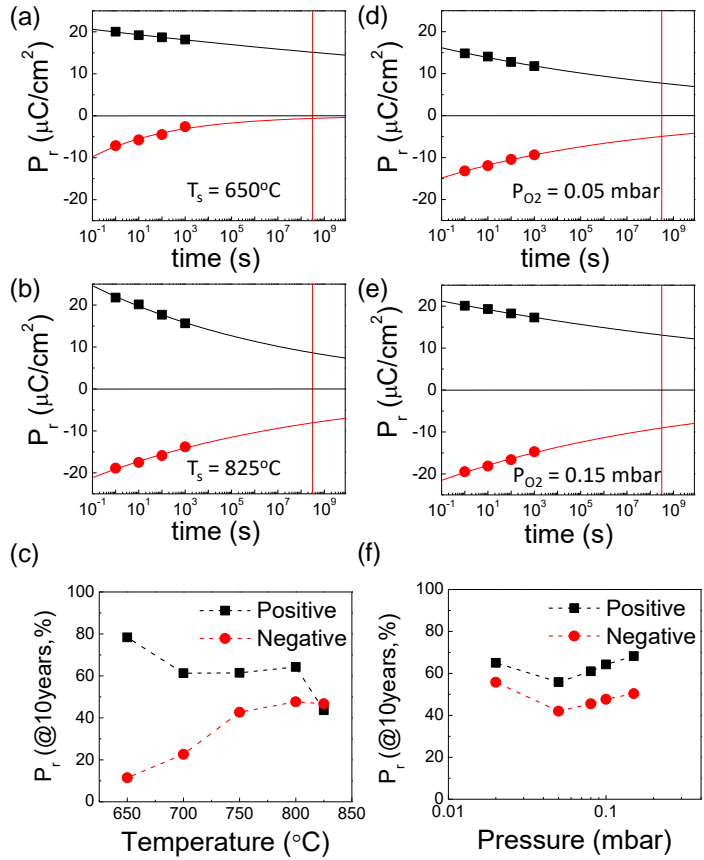

**Figure 2**: Retention of ferroelectric polarization poled positively (black squares) and negatively (red circles) for films deposited at $P_{O2}$ = 0.1 mbar and $T_s$ = 650 °C (a), 0.1 mbar and 825 °C (b), $P_{O2}$ = 0.05 mbar and $T_s$ = 800 °C (d), and 0.15 mbar and 800 °C (e). Vertical red line marks a time of 10 years. Normalized remanent polarization after 10 years, calculated from the $P_r = P_0 \, t_d^{-n}$ equation fitting, as a function of deposition temperature (c) and oxygen pressure (f).

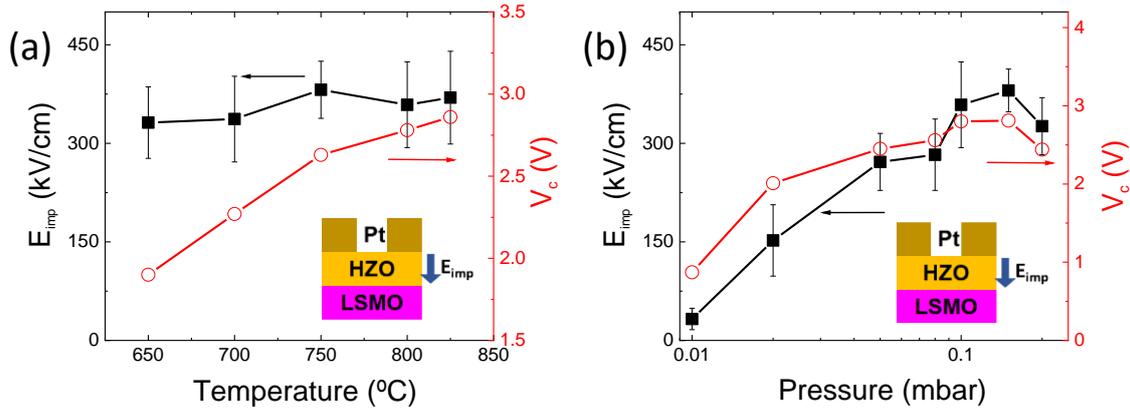

**Figure 3**: Dependence of imprint field and coercive field on the deposition temperature (a), and oxygen pressure (b). Sketches in the insets indicate the imprint field direction.

The dielectric constant $\varepsilon_r$ shows hysteresis loops in all $T_s$ series films (Figure 4a). The permittivity values are in the range of 28 to 36 (28 - 34 at saturation), with a lower permittivity for higher $T_s$. The oxygen pressure $P_{O2}$ (Figure 4b) has a greater influence. The film deposited at the lowest pressure, $P_{O2} = 0.01$ mbar, does not show significant dependence on the electric field, while the other films in the series show evident hysteresis. All films show negative horizontal shift in agreement with P-E loops (Figure 1). The sample deposited at 0.01 mbar has the smallest $\varepsilon_r$ around 16. As the oxygen pressure increases, the permittivity increases remarkably to around 30 at 0.1 mbar where it reaches a plateau when the oxygen pressure increases. Figure 4c shows the dielectric permittivity at saturation as a function of the deposition temperature and the oxygen pressure. Figure 4c shows that, in the investigated range of the growth window, the dielectric constant in the oxygen pressure series presents a greater variation than in the temperature series.

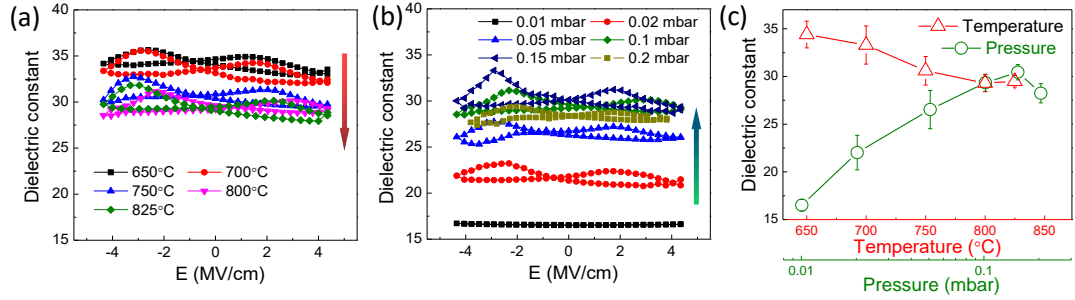

**Figure 4**: Dielectric permittivity versus applied electric field for films deposited varying growth temperature (a), and oxygen pressure (b). (c) Dielectric permittivity as a function of deposition temperature and oxygen pressure.

The wide range of the relative dielectric constant shown in Figure 4c could be mainly due to differences in the relative amount of HZO crystalline phases present in the films. Cubic, tetragonal, rhombohedral, orthorhombic and monoclinic polymorphs are reported in doped-$HfO_2$ films.[1,18-24] The dielectric constant varies depending greatly on the phase. Polycrystalline $Hf_{1-x}Zr_xO_2$ films typically present coexistence of monoclinic, tetragonal and orthorhombic phases, being the relative phases amount depending on the Zr content, thickness, and growth conditions.[1] The dielectric constant of these $Hf_{1-x}Zr_xO_2$ films is typically in the 20 - 40 range.[1,25] The dielectric constant of monoclinic and tetragonal phases, estimated theoretically for pure $ZrO_2$, is around 20 and 47, respectively.[26] In polycrystalline $Hf_{0.5}Zr_{0.5}O_2$ films, with high relative amount of orthorhombic phase, dielectric constant above 30 has been measured.[18] In the case of our epitaxial films, X-ray diffraction (XRD) and transmission electron microscopy (TEM) characterization revealed the existence of orthorhombic and monoclinic phases, but did not confirm presence of other phases (only $T_s$ = 800 °C and $P_{O2}$ = 0.1 mbar films were characterized by TEM).[6,7,27-29] XRD characterization revealed that the relative amount of orthorhombic and monoclinic phases is highly dependent on oxygen pressure during film growth, with an increased amount of orthorhombic phase as high $P_{O2}$ is.[6] Therefore, as the pressure increases, the increased amount of orthorhombic phase probably causes the large increase in the dielectric constant $\varepsilon_r$. This dependence is in agreement with the increase of remanent polarization of the films as $P_{O2}$ is higher.[6] On the other hand, the two films with dielectric constant higher than 30, which were deposited at lowest $T_s$ (650 and 700 °C), have smaller remanent polarization than the deposited at higher $T_s$. The high dielectric constant could be due to coexistence of tetragonal phase, not easily detected by XRD, although no signature of antiferroelectric behavior was observed in the dielectric and polarization hysteresis loops.

Fatigue in ferroelectric oxides is generally associated with oxygen vacancies that are generated by cycling, which are responsible for ferroelectric domain walls pinning, so some domains do not switch and the polarization reduces.[30] Charge injection is also suggested as mechanism of fatigue.[31] In polycrystalline hafnia, the defective regions would be located next to the interfaces with TiN electrodes[12,32] or at grain boundaries.[33,34] Defective regions would have low permittivity.[12] In the case of the epitaxial HZO films reported here, the top Pt electrodes used can be oxidized and scavenge oxygen from the HZO film. On the other hand, in these epitaxial HZO films on LSMO/STO(001), boundaries between crystal variants of the orthorhombic phase, and incoherent boundaries between monoclinic and orthorhombic grains,[29] are defective regions in which oxygen vacancies can accumulate. Monoclinic and orthorhombic phases present columnar topology, and the incoherent boundaries between both phases are perpendicular to the electrodes. Fatigue is more pronounced in epitaxial films deposited at high $T_s$ (Figure 1e) or high $P_{O2}$ (Figure 1f). The lower permittivity of the films grown at high $T_s$ could be due to oxygen defective grain boundaries, favoring the pinning of ferroelectric domains. In the $P_{O2}$ series, the low $P_{O2}$ films suffer less fatigue, while their relative permittivity is very low (about 22 in the $P_{O2}$ = 0.02 mbar film). The increased amount of monoclinic phase in the low $P_{O2}$ films reduces the overall permittivity of the film, likely hiding permittivity changes due to defects. The relative amount of monoclinic phase is higher in the low $P_{O2}$ and low $T_s$ films[6] and, however, they are less fatigued by repetitive field cycling (Figures 1c and 1f). Therefore, defects can have a more relevant role than the paraelectric phase on the ferroelectric fatigue. On the other hand, the out-of-plane lattice parameter was found to decrease monotonically with $T_s$ (from $d_{o-(111)}$ ~ 2.979 Å to ~ 2.959 Å) and $P_{O2}$ (from $d_{o-(111)}$ ~ 2.986 Å to ~ 2.954 Å).[6] Thus, the films with less fatigue has the longer $d_{o-(111)}$ parameters. However, epitaxial films (same thickness ~9 nm, $T_s$ = 800 °C, $P_{O2}$ = 0.1 mbar) on buffered Si(001) has highly contracted $d_{o-(111)}$ ~ 2.948 Å (due to the thermal mismatch with the Si(001) substrate), and they present very similar fatigue that the equivalent film on STO(001).[35] This discards a simple correlation between strain and resistance to fatigue.

Large remanent polarization is achieved in wide ranges of temperature (in the range of around 700 °C - 825 °C) and oxygen pressure (in the range of around 0.05 mbar - 0.2 mbar).[6] However, this wide growth window is not optimal considering fatigue and retention. For applications, high remanent polarization, low fatigue and long retention are properties that must be achieved simultaneously. Figure 5 summarizes the effects of $T_s$ and $P_{O2}$ on polarization, fatigue and retention. The requirement for all-properties optimized makes the growth window of epitaxial HZO films narrower, but remarkably wide enough ranges of $T_s$ and $P_{O2}$ allow films with optimal ferroelectric polarization, endurance and retention.

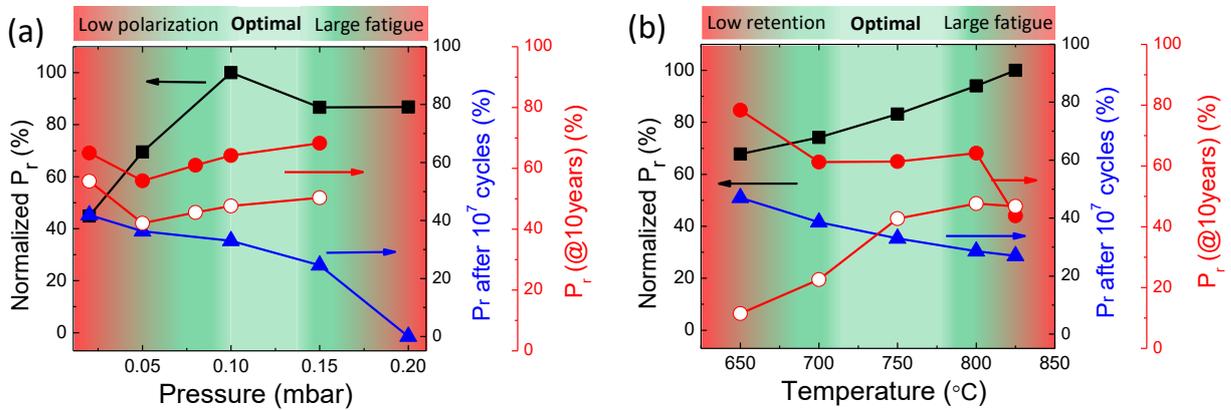

**Figure 5**: Dependences of the normalized remanent polarization in pristine state (black squares, left axis) and remanent polarization after $10^7$ cycles (blue triangles, right axis), and normalized remanent polarization after 10 years of positive (red solid circles, right axis) and negative (red empty circles, right axis) poling as a function of (a) oxygen pressure and (b) substrate temperature. Data of polarization in pristine state was reported in ref. 6. Colored areas summarize the main effect of the deposition parameters on polarization, fatigue and endurance.

In conclusion, the polarization retention and fatigue in the growth window (deposition temperature and oxygen pressure) of HZO epitaxial films have been determined. Films of thickness around 9 nm do not show significant wake-up effect, but exhibit pronounced fatigue that increases markedly with substrate temperature and pressure, while retention degrades in films deposited at low pressure. Defects located at grain boundaries are proposed to be critical, causing pinning of ferroelectric domains and an internal field. The dielectric permittivity varies significantly in the growth window, mainly varying oxygen pressure, as a consequence of different amount of defects or amount of paraelectric phase. The imprint field causes asymmetry in retention, and can severely degrade the polarization stability in films deposited at low temperature whose coercive field is smaller. However, there is long retention and low asymmetry over wide ranges of temperature and pressure. Nevertheless, optimal conditions to reduce fatigue produce less remanent polarization in the pristine state. The growth window of the epitaxial HZO is consequently narrower than when only remanent polarization is considered. But remarkably, films with all their critical properties, i.e. high remanent polarization, low fatigue and long retention, are achieved in sufficiently wide ranges of deposition temperature and pressure.

See the supplementary material for electrodes pictures and size measurements; current density versus voltage loops; and ferroelectric polarization loops after a few cycles.

Financial support from the Spanish Ministerio de Ciencia e Innovación, through the "Severo Ochoa" Programme for Centres of Excellence in R&D (SEV-2015-0496) and the MAT2017-85232-R (AEI/FEDER, EU), PID2019-107727RB-I00 (AEI/FEDER, EU), and MAT2015-73839-JIN projects, and from Generalitat de Catalunya (2017 SGR 1377) is acknowledged. IF acknowledges Ramón y Cajal contract RYC-2017-22531. JL was financially supported by China Scholarship Council (CSC) with No. 201506080019.

**Supplementary Material**

**S1**. Electrodes pictures from all samples reported in the present manuscript. Films deposited at $P_{O2}$ = 0.1 mbar and $T_s$ = 650 °C (a), 700 °C (b), 750 °C (c), 800 °C (d), and 850 °C (e), and at $T_s$ = 800 °C and $P_{O2}$ = 0.01 mbar (f), 0.02 mbar (g), 0.05 mbar (h), 0.1 mbar (i), 0.15 mbar (j) and 0.15 mbar (k).

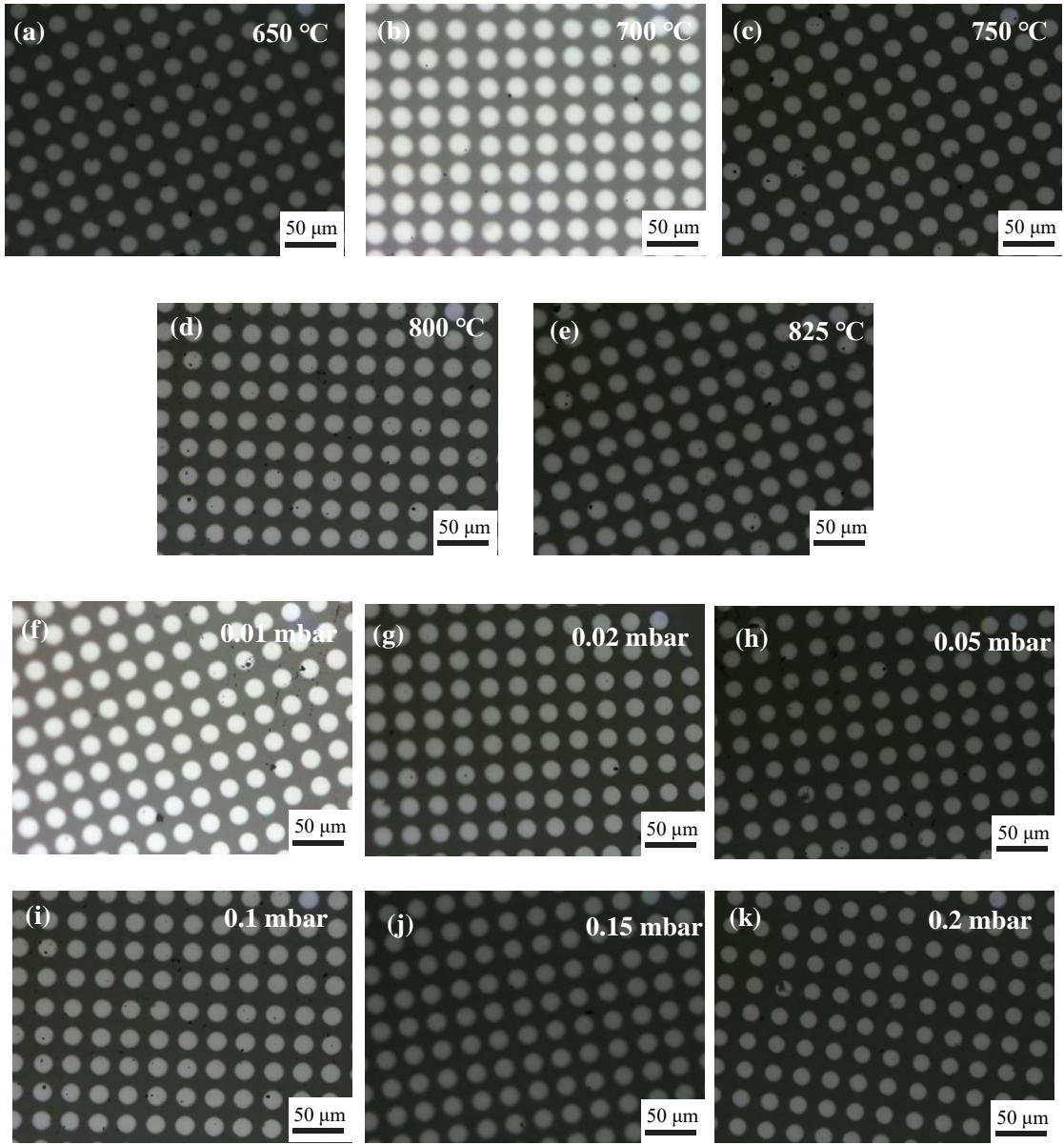

**S2**. Electrodes average diameter for samples set of samples growth at different $T_s$ (a) and $P_{O2}$ (b). Error bars are standard deviation of the diameter determined for all the electrodes shown in pictures of S1.

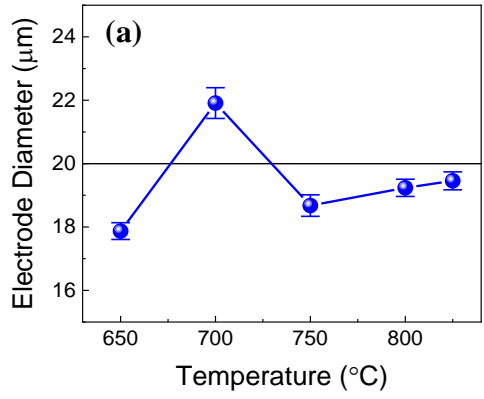 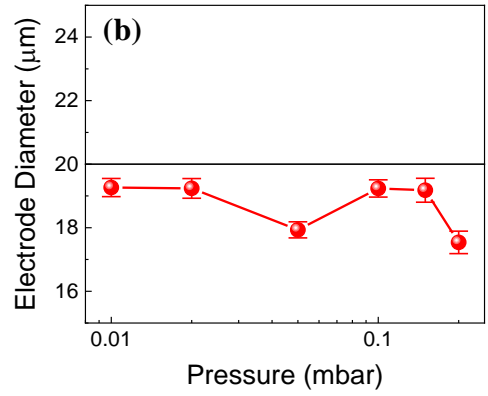

**S3**. J-V loops for the representative samples of Figure 1 are shown in panels (a,c,e,g). Zooms of the I-V characteristic at after $10^7$ cycles are shown in panels (b,d,f,h) for the corresponding samples. It can be observed that the ferroelectric current switching peaks (signaled by and arrow) gradually decrease while increasing the cycling number. It can be also observed a small decrease in the coercive field with cycling number. The ferroelectric current switching peaks at $10^7$ cycles are small but visible, being smaller for the samples grown at high $P_{O2}$ and high $T_s$. Extraction of the Pr value from ferroelectric current switching peaks integration (dashed region in b,d,f,h) gives a value of 1.22, 0.77, 0.40 and 0.27 μC/cm$^2$ respectively, which signals an overestimation of 2 μC/cm$^2$ in the direct extraction from P-E loops.

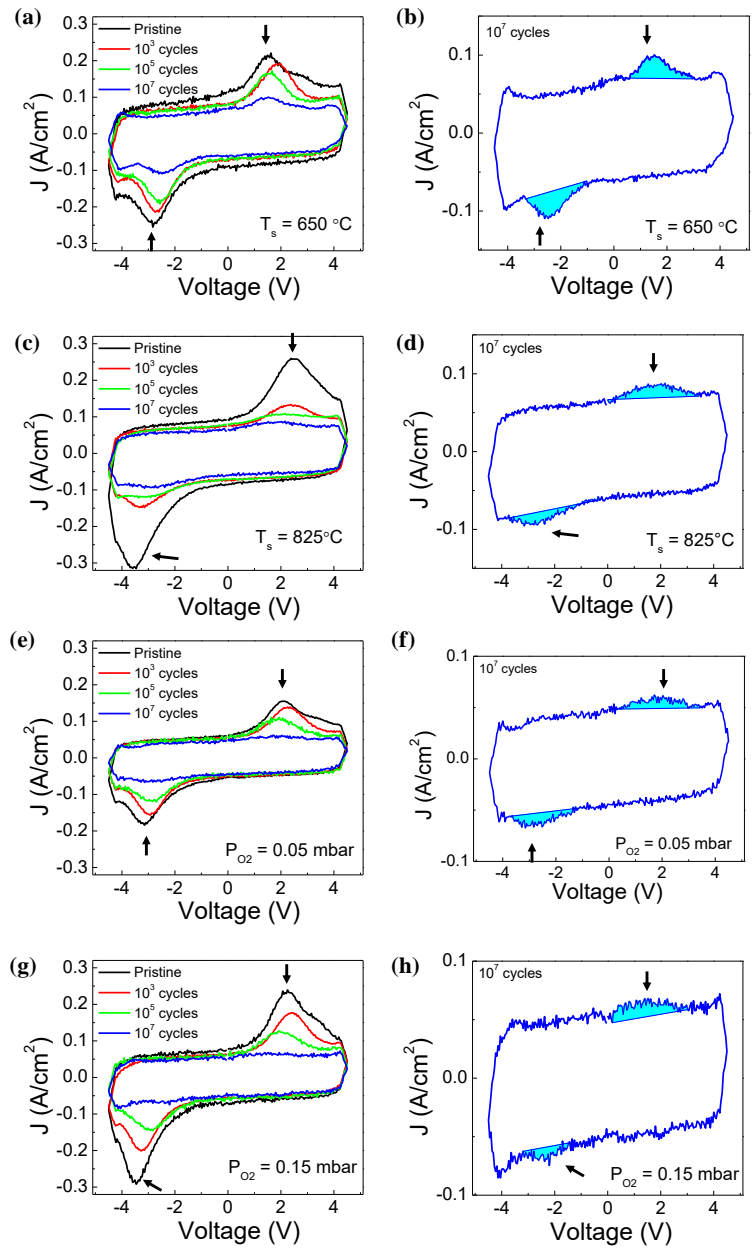

**S4**. Ferroelectric hysteresis loops, measured in pristine state and after 4 and 8 cycles at ~5.5 V, for films deposited at $P_{O2}$ = 0.1 mbar and $T_s$ = 650 °C (a), 700 °C (b), 750 °C (c), 800 °C (d), and 850 °C (e).

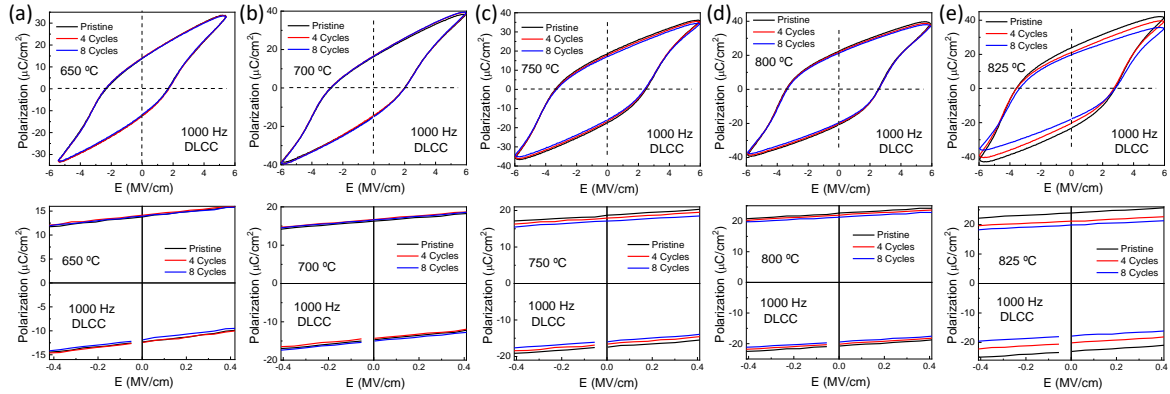

**S5**. Ferroelectric hysteresis loops, measured in pristine state and after 4 and 8 cycles at 4.5 V, for films deposited at $P_{O2}$ = 0.1 mbar and $T_s$ = 650 °C (a), 700 °C (b), 750 °C (c), 800 °C (d), and 850 °C (e).

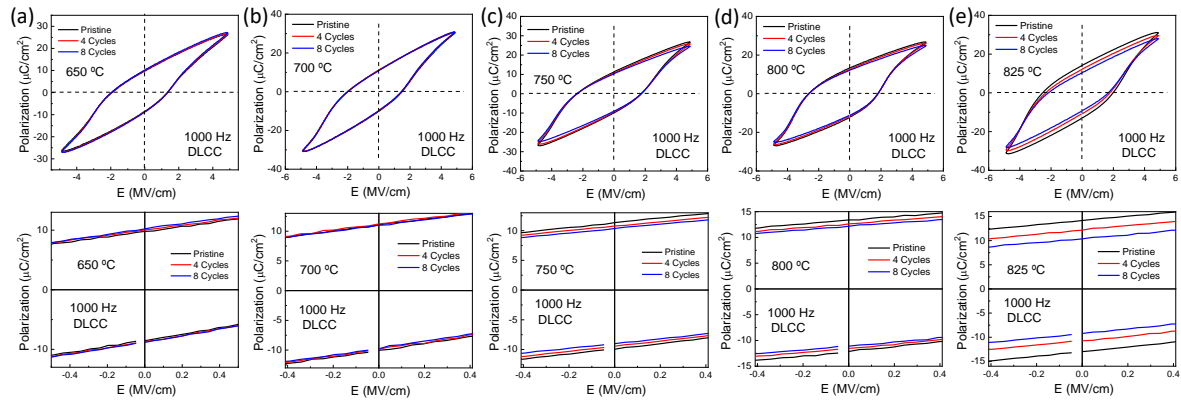

**S6**. Variation of remnant polarization (positive and negative average values) with number of cycles at ~5.5 V (a) and 4.5 V (b) of films deposited at $P_{O2}$ = 0.1 mbar and variable $T_s$.

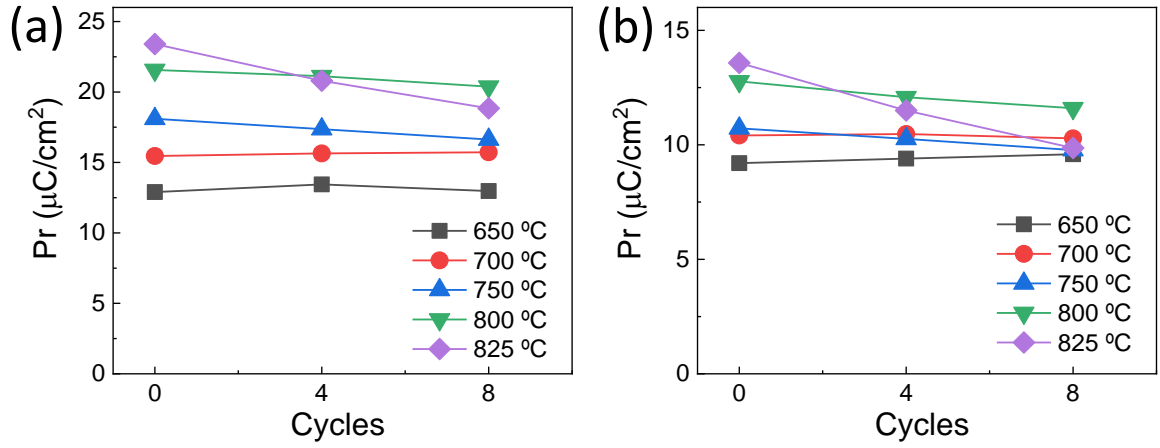

**S7**. Ferroelectric hysteresis loops, measured in pristine state and after 4 and 8 cycles at ~5.5 V, for films deposited at $T_s$ = 800 °C and $P_{O2}$ = 0.02 mbar (a), 0.05 mbar (b), 0.1 mbar (c), and 0.15 mbar (d).

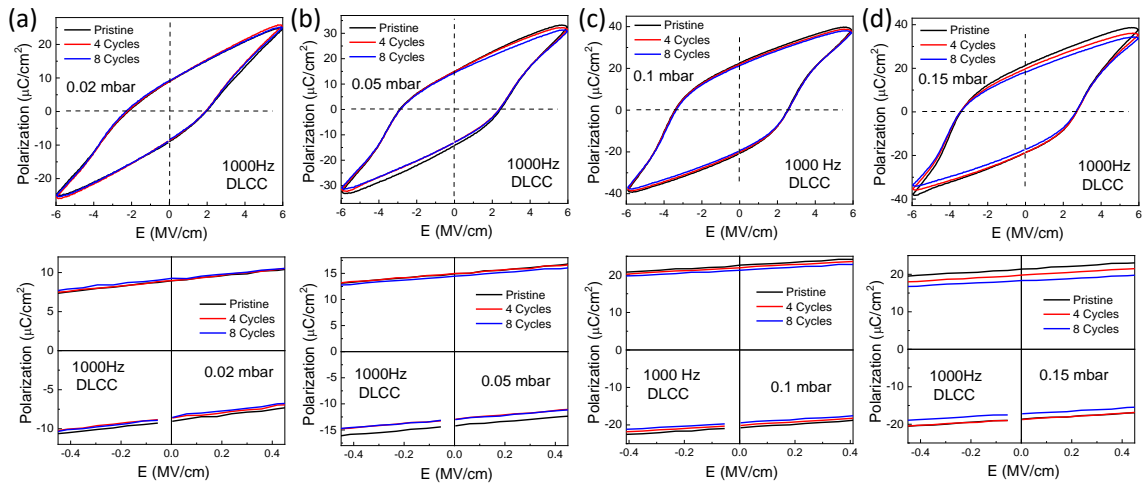

**S8**. Ferroelectric hysteresis loops, measured in pristine state and after 4 and 8 cycles at 4.5 V, for films deposited at $T_s$ = 800 °C and $P_{O2}$ = 0.02 mbar (a), 0.05 mbar (b), 0.1 mbar (c), and 0.15 mbar (d).

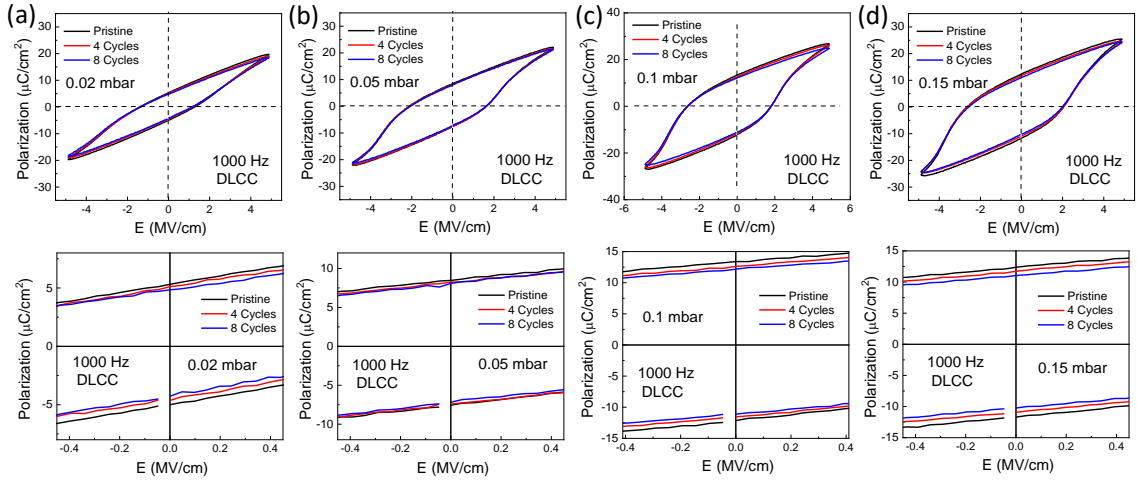

**S9**. Variation of remnant polarization (positive and negative average values) with number of cycles at ~5.5 V (a) and 4.5 V (b) of films deposited at $T_s$ = 800 °C and variable $P_{O2}$.

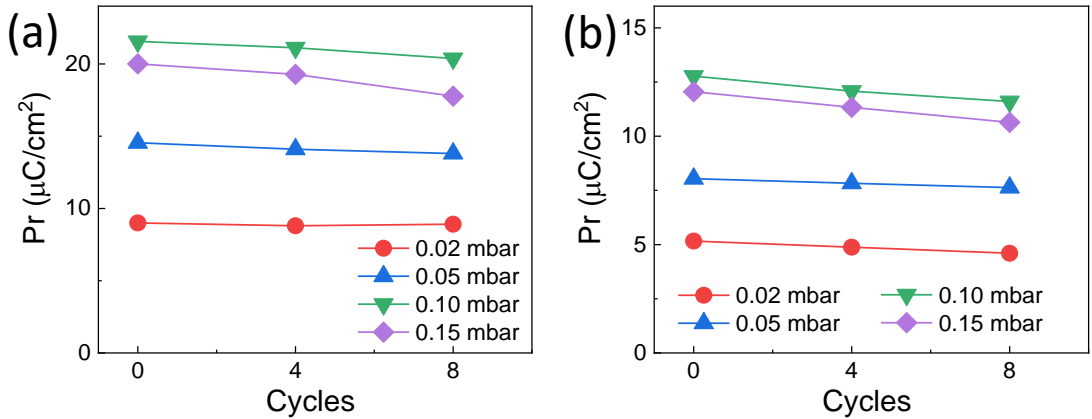